# The Effect of Uniform and Non-uniform Electron Density Models for Determining Shock Speed of a Type II Solar Radio Burst

D P S Nilagarathne[1*], J Adassuriya[2], H O Wijewardane[1]

[1]*Department of Physical Sciences, Rajarata University of Sri Lanka, Mihintale, Sri Lanka*
[2]*Arthur C Clarke Institute for Modern Technologies, Katubedda, Moratuwa, Sri Lanka*

*poornimanilagarathne@gmail.com

**ABSTRACT**

Solar flare is one of the most important solar activities which emit all electromagnetic waves in gigantic burst. The radio emission can be used to determine the physical properties of the solar flares. The e-CALLISTO worldwide network is designed to detect the radio emission of the solar flares and this study used the spectroscopic data from the e-CALLISTO system. Among the five types of solar radio bursts, this study was focused on type II radio bursts. The spectroscopic analysis estimated the shock speed of type II radio bursts using the uniform electron density model and the non-uniform electron density model of the sun. The shock speed is proportional to the electron density ($N_e$) and inversely proportional to the rate of change in electron density with altitude ($dN_e/dr$). The determined shock speed at the altitude of one solar radius is 2131 km/s for uniform model and 766 km/s for non-uniform model. Although the uniform electron density model is widely used we attempted the non-uniform electron density since in the active region of the sun, the electron densities are non-uniform. The estimated shock speeds of uniform density model is relatively high so that it is reasonable to use non-uniform electron density model for shock speed estimation of type II radio bursts.

## 1 INTRODUCTION

Solar radio bursts are structure of the frequencies that changes with time due to release of magnetic energy associated with sunspots. According to the radio astronomy, the frequencies of solar radio bursts vary from 70 MHz to 2.2 GHz. Yet most of the solar radio bursts occur in a low-frequency range such as 200 MHz [1][2]. Normally, this low frequency emission is considered to be a radio emission. The exact mechanism of type II burst is still a matter of debate; hence it was selected for the study. Shock speed or plasma velocity is one of the valuable parameter that gives much information about solar radio bursts. Although there are many work of shock speed of the type II solar burst calculated from uniform electron density model, where one considers the electron density around the sun to be constant, the shock speed calculation from non-uniform electron density model is an ongoing research. Therefore, the work describe here represents the results of the shock speed of type II radio burst compared with both uniform and non-uniform model.

## 2 METHODOLOGY

### 2.1 Shock Speed of Solar Radio Burst

The shock speed for uniform and non-uniform electron density models can be calculate by using the equation 1 as shown below.

$$V = \frac{2N \frac{df}{dt}}{f \frac{dN}{dr}} \tag{1}$$

where, *V* - shock speed, *N* - electron density, *f*- plasma frequency, $\frac{df}{dt}$ - drift rate, $\frac{dN}{dr}$ - change of electron density with respect to height

The type II radio bursts data that have been taken from CALLISTO system were used to carry this research. Data used here were taken from the published data on 16[th] of April 2014 fromlmsal.com[3]. The image taken by "ROSWELL-NM observatory; USA" of type II solar burst which was captured at 19:59:38 on 16/04/2014 was also used. The plasma velocities of the above mentioned radio bursts were calculated using uniform electron density model and non-uniform electron density model. Matlab software was used to analyze the spectroscopic data.

### 2.1.1 Analysis of Frequency

At the beginning of the analysis, FITS file of the flare image was read using Matlab command. For the removal of noises of the flare image, FITS file of a blank image which was captured few hours later was used. Noise removed image was cropped using Matlab software and the solar radio burst part was isolated. Afterward, the cropped isolated flare was loaded to an array defined [C, I]. Here, C is maximum value of each column of the matrix and "I" is the frequency channel. Then the frequency channel vs time graph was plotted to identify the distribution of the frequencies with respect to the time. Flare is spread in the time range of 700 to 2250 seconds. Once the actual frequency values of the image were shifted to new values due to cropping of the image, it was transformed to actual values by using a Python code. Generally, flare frequency is taken as the frequency of the starting point of the flare. For further calculations, initial frequency was taken as the flare frequency (f value) and $\frac{df}{dt}$ was calculated for the initial point by using Matlab software.

### 2.2 Uniform Electron Density Model

Newkirk model describes the electron density over chromosphere with an uniform electron density model approach [4,5]. Newkirk assumed that the variation of electron density in the quiet and active sun is uniform in the specific distance from the center of the sun. Electron density of the quiet sun is Newkirk × 1 and active sun is Newkirk × 20 [4]. The Newkirk equation is given by 2.

$$N_Q = N_0 \times 10^{4.32/R} \tag{2}$$

Where, $N_Q$ - electron density for quiet sun, $N_0$ - $4.2 \times 10^4$ , *R*- distance from the center of the sun in units of the solar radius ($R_0$)

Distance from the sun varied between 1.0 to 3.0 in 0.1 steps and electron density for different r values were obtained for the quiet sun and the active sun. $N_Q$ vs r graphs were plotted separately for quiet and active sun to observe the variation. Then $\frac{dN}{dr}$ values for different 'r' values were obtained from Matlab software. $N_Q$ vs r and $\frac{dN}{dr}$ vs r graphs for active and quiet sun are given in the Figures 2.1, 2.2, 2.3 and 2.4 respectively.

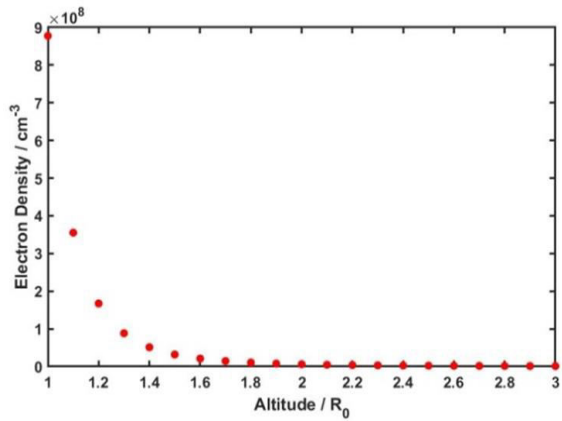 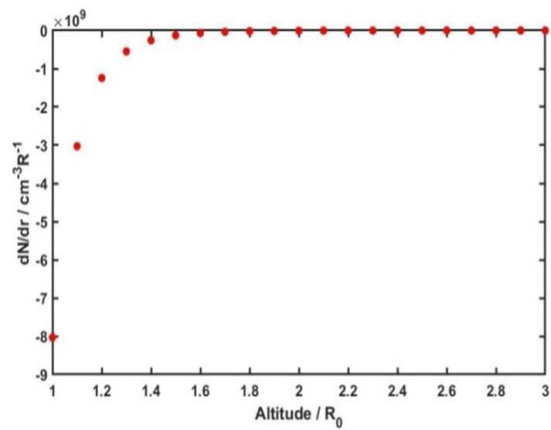

**Figure 2.1**: $N_Q$ vs r graph for active sun  **Figure 2.2**: $\frac{dN}{dr}$ vs r graph for active sun

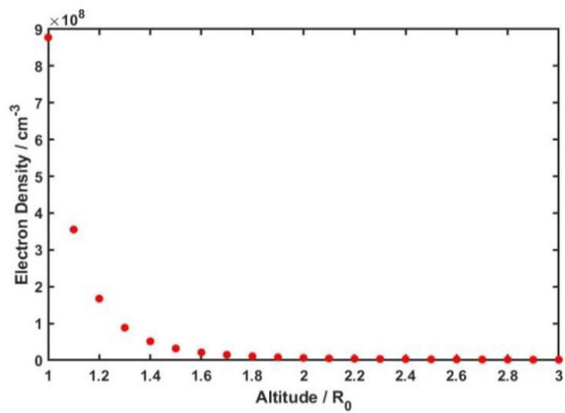 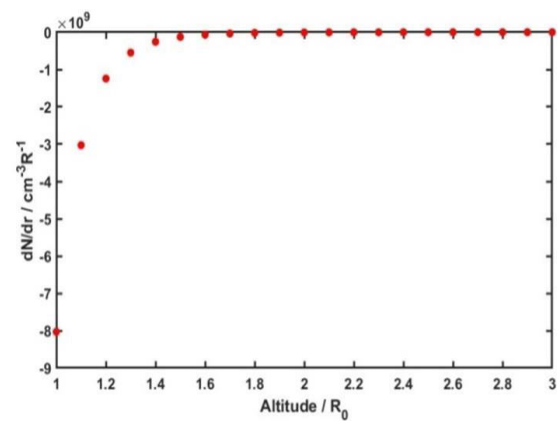

**Figure 2.3**: $N_Q$ vs r graph for quiet sun  **Figure 2.4**: $\frac{dN}{dr}$ vs r graph for quietsun

**2.2 Non-Uniform Electron Density Model**

In the non-uniform electron density model, the gradient of the electron density above the active region is steeper than the gradient of the quiet corona. So a new model was introduced to calculate the electron density above the active region of the sun when it is not uniform [5]. The model for the electron density at any spot in the corona is shown below.

$$N(R, \beta) = N_Q(R)\left[1 + C_1 e^{-\left(\frac{\beta^2}{2\sigma^2}\right)}\right] \qquad (3)$$

where, $N(R, \beta)$ - electron density at distance R from the solar center, $N_Q$ - Newkirk model, $\beta$ - chord between a point on the axis of the active region and distance R from the center of the sun, $C_1$ - a constant, $\sigma$ - a dispersion parameter

Here, $\beta^2 = (x - R\cos\theta)^2 + (y - R\sin\theta)^2 + z^2$ \qquad (4)

By the calculations done by Newkirk, the values for the parameters are adopted as $C_1 = 0.97$ and $\sigma = 0.235$. When consider the limb of the active region, $\theta$ becomes $90°$ and therefore equation 4 can be reduced to,

$$\beta^2 = x^2 + (y-R)^2 + z^2 \qquad (5)$$

To do the calculations for electron density of the non-uniform electron density model it is essential to find the point where the flare has started. To find the point, an image with a whole solar disk needed to be taken. The image of the solar disc is in the 2D plane. Therefore, it is necessary to convert the 3D equation to 2D equation in-order to find the electron density. When 3D equation is converted to 2D, coordinates of the X and Y axis remains and Z axis becomes 0. To calculate the electron density of the non-uniform electron density model, 2D image of a whole solar disc, which has been captured by e-CALLISTO spectrometer on 2014/04/16, was taken. To identify the area where the flare has occurred, FITS files of series of images were taken to identify the exact image that shows the variation of the sun at the moment that the flare was occurred. Fits file which has been published by sdo.gsfc.nasa.gov website was used to take the 2D images of the sun. The FITS files were read using IRAF software and the images were obtained. When FITS file was read, it was shown by a matrix which consists of 1024×1024 pixels. Then the maximum intensity of the sequence of images at time from 19:48 to 20:12 were taken. According to the data, graph was plotted with maximum intensity vs X-coordinates. By the graph, point 454 of the X-coordinate has the maximum intensity of 10723.1 and the coordinate of that point can be written as (454,452). Then by taking (454,452) as a reference point, further calculations done for different timeframes to confirm that the same point is the maximum. Then the intensities at the reference point (454,452) were taken. According to the data, graph was plotted with the intensity vs time to identify the variation of the pixel values at reference point. In the graph, the maximum intensity of 10723.1 is given at 20:00. To justify that the maximum intensity is given at 20:00, 5×5 error matrix for all time frames were calculated. Sum of all values in the error matrix for each time frames were taken as the intensity and further calculations were done. Maximum value of the sum of intensity is 119537.712 and it is found at 20:00. Image captured at 20:00 was taken as the exact image to calculate the coordinates. The point that the flare has occurred is identified as (454,452) by considering the origin as a corner point of the image. Yet for the calculation, the origin should be shifted to the middle of the solar disc, since the modified electron density model has been introduced by considering the origin from the middle of the solar disc. When consider the middle of the solar disc as the origin, the point that the flare occurred is available in the third quadrant of the Cartesian plane. Then, the values for both X-coordinate and Y-coordinate of that point become negative. If the image was loaded to the Matlab software, the left side upper corner is normally taken as the origin. When the origin was shifted to the center of the solar disc corresponding to the origin that was given by Matlab software, the point that the solar flare occurred was shifted to the fourth quadrant of the Cartesian plane. In order to clarify the error occurred by this change, the image was rotated in $270°$ degrees before the calculations were made (Figure 2.7). With the rotation, the point that the flare was occurred shifted back to the third quadrant. The new pixel coordinates of the origin was tabulated by Matlab and they are (523.307831102875, 500.564806364643). Hence the real coordinates of the point was calculated by subtracting the value (454,452) taken by IRAF software. Therefore the real coordinate of the point that the flare was occurred is given by (-69.3078311028750, -48.5648063646430). To find the β value, X and Y coordinates should be transferred to the unit of solar radius. This was done by checking the ratio of the radius of the solar disc to the actual solar radius. After that by converting the amount of the solar radius denoted by a single pixel value and multiplying that with the X and Y values new coordinates were obtained in the unit of solar radius. Accordingly, β value was calculated by using equation 5. Then electron density values for the different 'r' values were obtained.

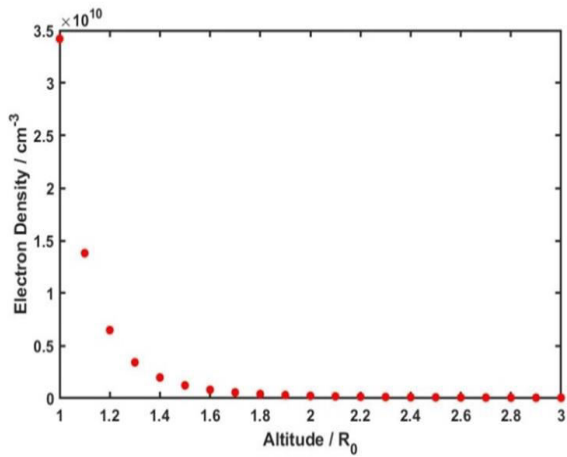 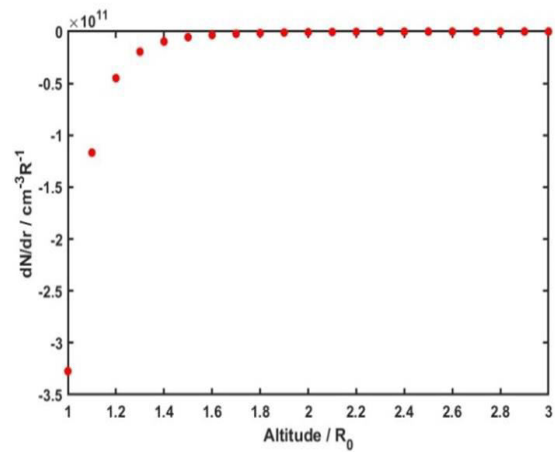

**Figure 2.5:** $N_Q$ vs r graph  **Figure 2.6:** $\frac{dN}{dr}$ vs r graph

The values obtained for $N$ and $\frac{dN}{dr}$ with the non-uniform density model were used in the calculation of shock speed. Finally obtain the flare occurred point as shown in Figure 2.8.

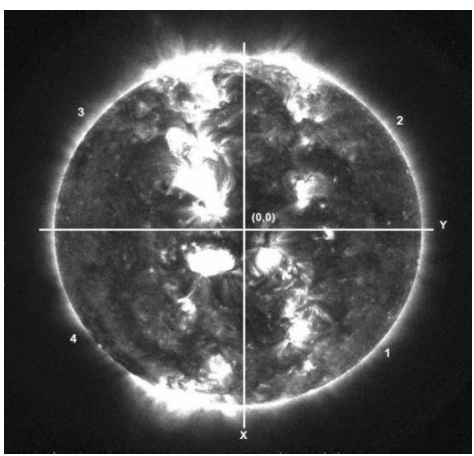 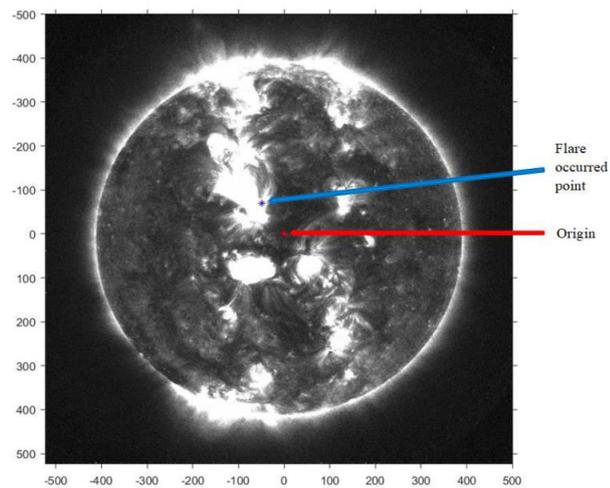

**Figure 2.7:** $270°$ degrees rotated image point

**Figure 2.8:** Origin and the flare occurred

## 3 RESULTS AND DISCUSSION

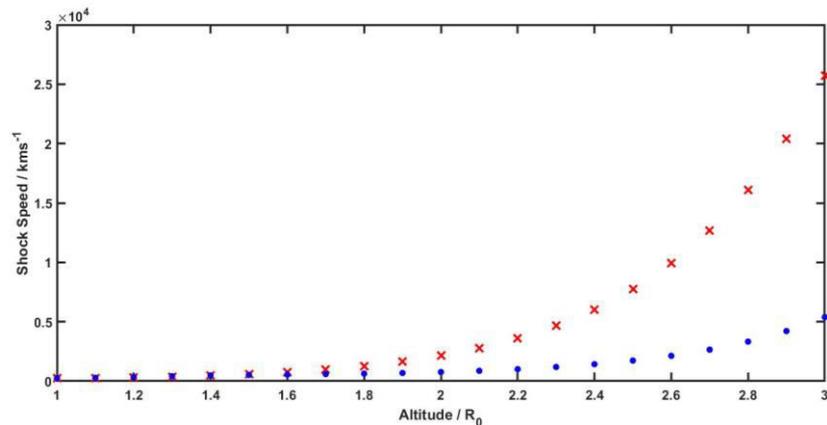

**Figure 3.1:** Velocity curves of uniform and non-uniform electron density models

The plot of velocities vs r tabulated from the uniform and non-uniform electron density models shown in the Figure 3.1, demonstrate a slight difference in its curves. According to the equation 2 and 3, the electron density of the uniform electron density model is lower than that of the electron density of the non-uniform electron density model in a given distance. Because of the presence of active regions the electron density of the non-uniform electron density model is higher than the uniform model. The rate of increase in shock speed with the altitude in non-uniform electron density model is less than that of the uniform electron density model. The shock speed is proportional to the electron density ($N_e$) and inversely proportional to the rate of change in electron density with altitude ($dN_e/dr$). Therefore the shock speed for uniform electron density is quite higher than that of the non-uniform model. The change of the velocity starts from the 2.0 $R_0$ distance. In that point the shock speed of the uniform model is 2131 km/s and in non-uniform model it is 766 km/s. Although there are no clear evidences for the speed for the type II bursts using non-uniform density model, literature says that the typical shock speed is less than 1000 km/s in the above distance[6]. Therefore the value we obtained from the non-uniform electron density model is quite closer to the typical value. Normally, electron density of the active region is not uniform and there are many researches done with considering the electron density in an active region is uniform. This research is mainly focused on the comparison of the uniform electron density model with a non-uniform electron density model. Hence it was done with a unique dataset. Since the result is more closer to given values in the literature it is more considerable to take non-uniform electron density model to calculate the electron density of an active region. Nevertheless if one need to clarify the actual value and make an exact decision, then it is more appropriate to check with more sets of data obtained from e-CALLISTO global network. Therefore in future same type of calculations will be done for different sets of data in order to decide whether the study of shock speed of type II bursts can be done more accurately with the non-uniform electron density model by Newkirk.